# Magnetic Proximity Effect in a van der Waals Moiré Superlattice

Qingjun Tong[1,2,*], Mingxing Chen[3], Wang Yao[2]

[1]*School of Physics and Electronics, Hunan University, Changsha 410082, China*
[2]*Department of Physics and Center of Theoretical and Computational Physics, University of Hong Kong, Hong Kong, China*
[3] *Key Laboratory of Low-Dimensional Quantum Structures and Quantum control of Ministry of Education, Synergetic Innovation Center for Quantum Effects and Applications (SICQEA), Hunan Normal University, Changsha 410081, China*

**Abstract:** We investigate the magnetic proximity effect in van der Waals heterostructure formed by a monolayer semiconductor stacked on a 2D ferromagnet, where the lattice mismatch and twisting between the layers typically lead to the formation of moiré pattern. We find that the magnetic proximity effect arising from the spin dependent interlayer coupling depends sensitively on the interlayer atomic registry. Consequently, in the moiré pattern, the spatial variation of the atomic registry leads to a lateral modulation of magnetic proximity field. Such moiré modulated magnetic proximity effect manifests as a miniband spin splitting that strongly depends on the moiré periodicity which can be mechanically tuned by a relative twisting and/or strain between the layers. We also show, because of the moiré modulation on the interlayer distance, a perpendicular electric field can be used to control the miniband spin splitting. Our results suggest potential nanodevices where the moiré modulated magnetic proximity effect can lead to unique spin controllability.

## I. Introduction

The recent discovery of long-range magnetic order in atomic thin crystals opens new opportunities for the investigation of 2D magnetism and for low-power spintronic applications [1-3]. The layered 2D magnets feature a remarkable controllability in the interlayer magnetic order by external electric or magnetic fields [4-8]. The layered antiferromagnetic order in some 2D magnets is also utilized to realize magnetic tunnel junction in the atomically thin limit, with giant tunneling magnetoresistance values demonstrated [9-14].

Van der Waals (vdW) heterostructures provides a powerful approach towards engineering functional materials and novel platform to explore fundamental physics [15]. A highly interesting aspect of vdW heterostructure is the general presence of moiré pattern due to the lattice mismatch and/or twisting in the stacking [16-20]. In a long-period moiré pattern the atomic registry in each local resembles lattice-matched commensurate bilayer but changes local-to-local over long range. Generally the atomic registry determines the interlayer interaction [21-23], as a result, the local physical property varies periodically in the moiré superlattice. This idea has led to the exciting possibilities to engineer topological moiré minibands [24,25], topological insulator superstructures [26], moiré excitons [27-31], and moiré defined magnetic textures and skyrmions [32].

When integrated with other 2D materials to create vdW heterostructures, layered magnetic materials can provide an efficient platform for spin control in the nonmagnetic layer via the magnetic proximity effect [33-41]. Recent experiments have observed control of the spin and valley pseudospin in monolayer TMDs stacked on a thin ferromagnet [37,38]. The unique gate tunability of magnetic order in the 2D magnets further suggests remarkable control on the magnetic proximity effect [5,6,14]. The magnetic proximity effect in the context of a moiré pattern can be a highly interesting issue. With the general dependence of interlayer physics on the stacking in vdW heterostructures [26,32,42], it can be expected that the magnetic proximity field as an interface effect would depend on the atomic registry as well. This implies new opportunities to exploit the magnetic proximity effect in the vdW heterostructures.

In this work, we investigate the magnetic proximity effect in a long-period moiré superlattice formed by a magnet monolayer ($CrI_3$) and a semiconductor monolayer (BAs). In lattice-matched heterobilayer configurations, our first-principles calculations show a strong dependence of the magnetic proximity field on the interlayer atomic registry, with the induced Zeeman splitting varying between 0 and 12.7meV. As a result, in the moiré pattern, where the atomic registry changes smoothly, there is a strong lateral modulation in the magnetic proximity field. Such magnetic proximity effect in the moiré superlattice manifests as the spin splitting in the moiré miniband, which strongly depends on the moiré periodicity that can be tuned by a relative twisting and/or strain between the layers. We also show remarkable electric tunability of the miniband spin splitting by a perpendicular electric field, exploiting the moiré modulation in the interlayer distance. We find that an electric field can eventually lead to the spatial separation of the spin-up and spin-down wave functions in the moiré superlattice. The well-localized spin-polarized state in the miniband suggests that vdW moiré can be utilized for engineering programmable spin-polarized quantum dot arrays, which can be further tuned into waveguide arrays by a uniaxial strain.

## II. Moiré pattern in vdW heterobilayer: system description

We consider the example of a heterobilayer formed by monolayer semiconductor BAs and ferromagnet $CrI_3$, both of which has a hexagonal structure. Our result is general and can be extended to other similar systems. The lattice constant $a$ of BAs and $CrI_3$ is 3.39 Å and 6.867 Å respectively, with the lattice mismatch between $1 \times 1$ unit cell of $CrI_3$ and $2 \times 2$ supercell of BAs being ~1.3% [43,44]. This small lattice mismatch leads to a long-period moiré superlattice, with its periodicity can be further tuned by a relative twisting and/or strain between the layers. For heterobilayer with small lattice mismatch $\delta$ and/or twisting angle $\theta$, the moiré periodicity $b \approx a/\sqrt{\delta^2 + \theta^2}$. In a long-period moiré, each local region $\mathbf{R}$ can be regarded as a commensurate bilayer with identical lattice constants, while in different locals the interlayer atomic registry differs (c.f. Fig. 1(a)). For lattice-matched bilayers, the interlayer atomic registry can be characterized by an interlayer translation vector $\mathbf{r}$ defined in Fig. 1(b). Note that $\mathbf{r}$ is only well defined in a $1 \times 1$ unit cell of BAs due to the translation invariance in lattice-matched bilayers. Therefore the moiré superlattice can be described by a set of lattice-matched bilayers parameterized by a vector **r** characterizing the local-to-local variation of interlayer atomic registry over long-range in the moiré.

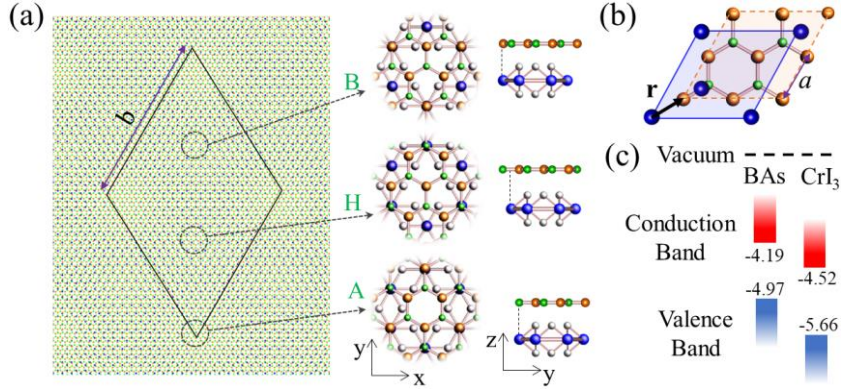

FIG. 1. (a) Moiré pattern formed in BAs/CrI$_3$ heterobilayer. The atomic structures in the three locals are magnified showing lattice-matched bilayer configurations, named as A, B, and H respectively. A side view is given on the right panel. The blue and gray spheres are Cr and I atoms of FM monolayer. The orange and green spheres are As and B atoms of the monolayer semiconductor. (b) An interlayer translation vector **r** can be used to describe the locally different atomic registries in the moiré. (c) Band alignment of BAs and CrI$_3$ relative to the vacuum level showing a type-II configuration. Energy is in unit of eV.

First-principles calculation shows that the heterobilayer BAs/CrI$_3$ has type-II band alignment with conduction band contributed from CrI$_3$ layer and valence band from BAs layer (c.f. Fig. 1(c)). Because the conduction band contributed from BAs layer is covered by the one from CrI$_3$ layer in a heterobilayer, here we are interested in the valence band edge of BAs layer located around K point in the first Brillouin zone. Explicitly we will focus on the spin splitting at valence band edge of BAs arising from the magnetic proximity field from CrI$_3$.

For systems with small spin-orbital coupling, the major contribution to the magnetic proximity effect is from the spin-conserved interlayer hopping. With the large spin splitting in the energy band in CrI$_3$, the interlayer hopping between BAs and CrI$_3$ then results in a spin dependent second order energy correction to shift the BAs energy band. As a result, the spin degeneracy in BAs is lifted, which is a magnetic proximity effect imprinted from the monolayer ferromagnet. This proximity field induced spin splitting is $E_s \approx \sum_i \left(\frac{1}{E_i} - \frac{1}{E_i + \Delta_i}\right) t_i^2$, where $E_i$ is the energy separation between the valence band of BAs and the $i$-th band of CrI$_3$ with spin splitting energy $\Delta_i$ and $t_i$ is the interlayer hopping amplitude between them. First-principles results show that the valence band of BAs is ~450meV bellow the conduction band of CrI$_3$ and ~700meV above its valence band. Therefore, interlayer hopping from the valence band of BAs to the conduction band of CrI$_3$ has more prominent contribution to the proximity effect than to its valence band.

In the following, we establish a quantitative model for the magnetic proximity effect through the first-principles calculations of the heterobilayer band structures. Direct calculation of a long-period moiré can be challenging because of the large number of atoms involved. We first examine lattice-matched heterobilayers of various stacking registries, and the dependence of the magnetic proximity field on the registries. We then establish a model to account for the magnetic proximity effect in a long-period moiré under the local approximation.

### III. Magnetic proximity effect in lattice-matched heterobilayers

The magnetic proximity effect is related to the interlayer coupling, which depends on the atomic orbits and interlayer atomic registry of the bilayer. In lattice-matched commensurate case, the hopping matrix element between valence band of semiconductor BAs layer and the energetically nearby $i$-th band of ferromagnetic CrI$_3$ layer is $t_i = \langle \psi_v^S(r) | \hat{H} | \psi_i^F(r) \rangle$, where $\hat{H}$ is interlayer interacting Hamiltonian. $\psi_i^{S(F)}(r)$ denotes the Bloch state at K point in the semiconductor (ferromagnetic) monolayer. With the atomic orbits fixed, interlayer atomic registry determines solely the interlayer coupling. In the following we first discuss the magnetic proximity effect for $\hat{C}_3$ rotation symmetric bilayers, in which the interlayer coupling can be understood from symmetry analysis. Then we extent our results to bilayers with general interlayer atomic registry. The results are also compared with first-principles calculations.

### A. Magnetic proximity effect in bilayers with $\hat{C}_3$ rotation symmetry

The three different high-symmetry configurations are shown in the insets in Fig. 1(a), corresponding respectively to the cases that an As atom of the semiconductor layer sits right on top of an A or B magnetic atom or hollow center in the ferromagnetic monolayer. In the following we name these three configurations as A, B, and H respectively. The corresponding translation vectors are $\mathbf{r} = \{0, -\frac{\mathbf{a}_1+\mathbf{a}_2}{3}, \frac{\mathbf{a}_1+\mathbf{a}_2}{3}\}$ with $\mathbf{a}_{1,2}$ being the two unit vectors of monolayer BAs. At these configurations, the interlayer Hamiltonian is invariant under $\hat{C}_3$ rotation. Such an operation on Bloch states $\psi_m^{S(F)}(r)$ at K point leads to

$$\hat{C}_3 \psi_m^{S(F)}(r) = C_3^{S(F)} \psi_m^{S(F)}(r),$$

where $C_3^{S(F)} = e^{i\phi^{S(F)}(X)} \gamma_m^{S(F)}$ and $\gamma_m^{S(F)} = e^{-i\frac{2m\pi}{3}}$ is the eigenvalue of $\hat{C}_3$ operating on the atomic orbit of the semiconductor (ferromagnetic) layer with $m$ being its quantum number. The phase $\phi^{S(F)}(X)$ arises from the plane-wave component in the Bloch function depending explicitly on the rotation center $X = \{A, B, H\}$. Then the interlayer hopping matrix element is $t_{vi} = \langle \hat{C}_3 \psi_v^S | \hat{H} | \hat{C}_3 \psi_i^F \rangle = C_3^{S*} C_3^F t_{vi}$, therefore a non-zero interlayer coupling requires that $C_3^{S*}C_3^F = 1$.

TABLE I. The eigenvalue of $\hat{C}_3$ operating on the Bloch state at K point of CrI$_3$ for different rotation center X = {A, B, H}. $c$ and $c_{1,2}$ are band indices stand for conduction band and two energy bands above it (c.f. Fig. 9 in Appendix B).

|  | $c$ | $c_1$ | $c_2$ |
|---|---|---|---|
| $C_3^F(A)$ | $e^{i2\pi/3}$ | $e^{-i2\pi/3}$ | $e^{-i2\pi/3}$ |
| $C_3^F(B)$ | $e^{-i2\pi/3}$ | 1 | 1 |
| $C_3^F(H)$ | 1 | $e^{i2\pi/3}$ | $e^{i2\pi/3}$ |

The eigenvalue $C_3^F(X)$ for different energy bands and rotation centers of ferromagnetic CrI$_3$ are given in Table I. The eigenvalue $C_3^S = 1$ for the valence band of BAs when the rotation center is fixed at an As atom. Then the configurations with allowed interlayer hopping from the valence band of BAs to the three energetically nearby bands of CrI$_3$ are those with eigenvalue of 1 listed in Table I. Explicitly for A configuration, the interlayer hopping is forbidden. For B and H configurations, the hopping to the conduction band or bands above it is allowed. Because these energy bands are spin-polarized, their interaction with the valence band of BAs will lead to the spin splitting in the latter, which is the magnetic proximity effect. Furthermore, the conduction band is energetically closer to the valence band of BAs than the other two bands, so we expect that the spin splitting in H configuration is more prominent than the one in B configuration.

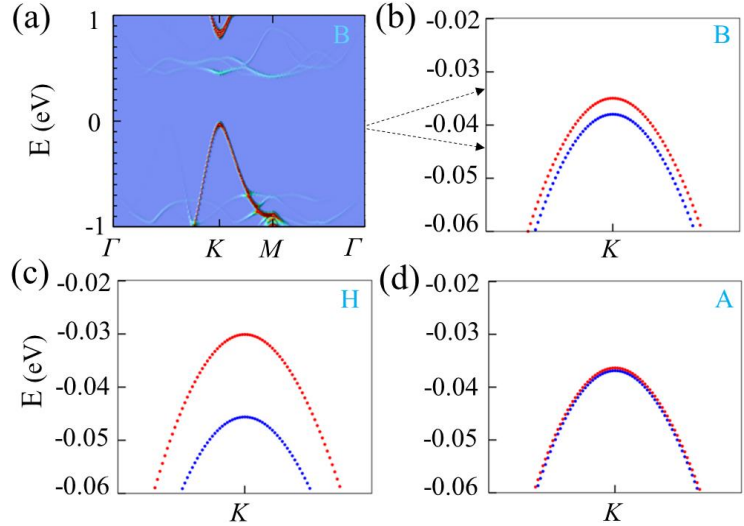

FIG. 2. First-principles band structures of the $\hat{C}_3$-rotation symmetric lattice-matched heterobilayer BAs/CrI$_3$. (a) Band structure for B configuration projected on the BAs layer. Two spin resolved (red and blue dots) energy bands around the valence band at K point for B (b), H (c), and A (d) configurations. A 1.3% strain is applied to a 2x2 superlattice of BAs layer to obtain a commensurate structure with 1x1 unit cell of CrI$_3$ Layer.

First-principles calculations conform to our results based on symmetry analysis. The heterobilayer shows a type-II band alignment with the valence band being mainly contributed from the semiconductor BAs layer (c.f. Fig. 2(a)). Figs. 2(b)-2(d) show the spin-resolved band structure around the valence band edge for the three symmetric configurations. The spin splitting at K point in H configurations is largest $E_s^H \approx$ 12.7meV. The one in B configuration is $E_s^B \approx$ 2.6meV. For A configuration the spin splitting is negligible.

## B. Magnetic proximity effect in general cases

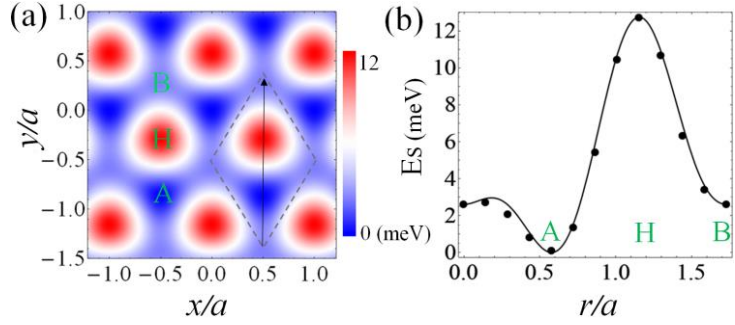

FIG. 3. (a) Interlayer translation dependence of spin splitting $E_s$. (b) Comparison of the result of Eq. (1) (black line) with first-principles result (black dots) along the black-solid line in (a). A, B, and H mark the symmetric configuration points.

For bilayers with a general $\mathbf{r}$, $\hat{C}_3$ rotation symmetry is broken. The interlayer translation $\mathbf{r}$ dependence of spin splitting at the valence band edge in the semiconductor BAs layer can be formulated as [27],

$$E_s(\mathbf{r}) \cong E_s^H \left| e^{-i\mathbf{K}\cdot\mathbf{r}} + e^{-i\hat{C}_3\mathbf{K}\cdot\mathbf{r}-\frac{2\pi}{3}} + e^{-i\hat{C}_3^2\mathbf{K}\cdot\mathbf{r}+\frac{2\pi}{3}} \right|^2 + E_s^B \left| e^{-i\mathbf{K}\cdot\mathbf{r}} + e^{-i\hat{C}_3\mathbf{K}\cdot\mathbf{r}+\frac{2\pi}{3}} + e^{-i\hat{C}_3^2\mathbf{K}\cdot\mathbf{r}-\frac{2\pi}{3}} \right|^2 \quad (1)$$

where $\mathbf{K}$ is the wavevector for the K corner of the first Brillouin zone. The two terms are contributions of hopping from the conduction band and the ones above it in the ferromagnetic CrI$_3$ layer. The amplitude $E_s^H = 12.7$meV ($E_s^B = 2.6$meV) is the spin splitting at H (B) configuration obtained from Fig. 2. The dependence of $E_s(\mathbf{r})$ on $\mathbf{r}$ is shown in Fig. 3(a). We also compared the above approximate result with first-principles result, which shows a good agreement (c.f. Fig. 3b). These analytical and numerical results show that the magnetic proximity effect depends sensitively on the atomic registry in vdW heterostructures.

## C. Interlayer atomic registry dependence of band-edge energy

The magnetic proximity effect only accounts for the difference in the interlayer interaction of the two spin components reflected in the spin splitting at the valence band edge. The vdW interlayer interaction also leads to a global shift of the band-edge energies relative to the vacuum level for both spins. This shift also depends on the interlayer atomic registry. The contributions of these two effects determine the actual position of the band edge of the heterobilayer.

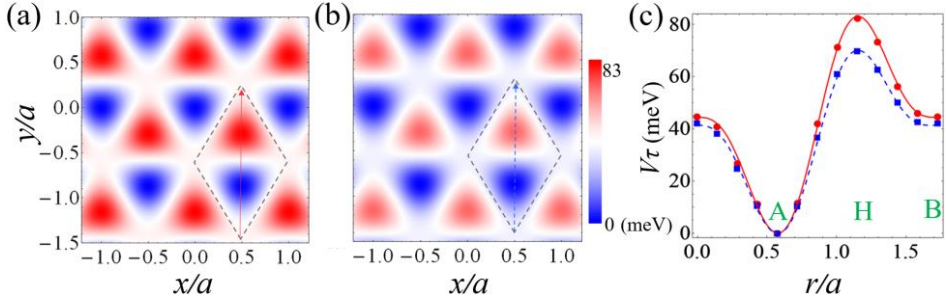

FIG. 4. Interlayer translation dependence of the valence band-edge energy for spin-up (a) and spin-down (b) components. (c) Comparison of the result of Eq. (2) (red-solid and blue-dashed lines) with first-principles result (red dots and blue squares) along the lines in (a) and (b).

The interlayer atomic registry dependence of band-edge energy can be accurately approximated by a Fourier expansion that includes only the three reciprocal lattice vectors in the first shell. Because of the $\hat{C}_3$ rotation symmetry, the energy can be written as [28]

$$V_\tau(\mathbf{r}) = v_\tau \sum_{i=1,2,3} \cos(\mathbf{G}_i \cdot \mathbf{r} + \varphi_\tau) \quad (2)$$

$\mathbf{G}_i$ is reciprocal lattice vector with $\mathbf{G}_1 = (0, \frac{4\pi}{\sqrt{3}a})$ and $\mathbf{G}_3 = \hat{C}_3 \mathbf{G}_2 = \hat{C}_3^2 \mathbf{G}_1$. The parameters $\{v_\tau, \varphi_\tau\} \approx \{83\text{meV}, 152.3°\}$ for spin-up $\tau = \uparrow$ and $\{70\text{meV}, 156°\}$ for spin-down $\tau = \downarrow$ components can be fitted from first-principles results at high-symmetry configurations. Note that we have chosen the minimum of band-edge energy as zero. Figs. 4(a) and 4(b) show the interlayer atomic registry dependence of band-edge energy for the two spin components. Along the lines, the analytical result is compared with first-principles result showing a good agreement (c.f. Fig. 4(c)). The difference between the band-edge energies of the two spins is just the magnetic proximity effect shown in Fig. 3.

### IV. Magnetic proximity effect in vdW moiré

Armed with the results in lattice-matched commensurate cases, we now turn to study the magnetic proximity effect in vdW moiré. In a moiré pattern, because the atomic registry changes local to local, it is infeasible to directly define the magnetic proximity effect from the local commensurate case. However the moiré periodicity guarantees the application of Bloch theory and gives rise to the moiré miniband, in which the spin splitting at miniband edge then defines the magnetic proximity effect in a moiré.

### A. Spin-polarized miniband

The interlayer atomic registry dependence of the band-edge energy in lattice-matched commensurate bilayers suggests that in moiré pattern of an incommensurate heterobilayer, where the atomic registry changes smoothly, there is a lateral modulation of band-edge energy. This modulation can be regarded as a confinement potential leading to the localization of band-edge state.

To clarify this modulation effect, it is convenient to define a moiré potential under the local approximation [45]. In a long-period moiré, the local atomic registry changes smoothly, which is characterized by the translation vector **r** taking continuously the values in a unit cell. We assume that the moiré is formed between two rigid lattices (i.e. no reconstruction), where the mapping between the local registry **r** and location **R** is linear. Therefore the **r** dependence of the interlayer potential in commensurate bilayers can be directly mapped to the location (**R**) dependence of the interlayer potential in a moiré [26,27,32]. Under this approximation, the moiré modulated interlayer potential as a function of **R** is isomorphic to $V_\tau(\mathbf{r})$ that we have formulated in Eq. (2), i.e. $V_\tau(\mathbf{R}) = V_\tau(\mathbf{r}(\mathbf{R}))$. This defines a moiré potential accounting for the influence of CrI$_3$ layer for the valence band of BAs layer. Note that this potential is spin-dependent arising from the magnetic proximity effect. When the in-plain relaxation is considered, the equilibrium atomic structure should be determined by the competition between adhesion and monolayer's elastic properties. This would change quantatively the profiles of spin splitting and band edge energy in the moiré pattern. However the general picture of moiré modulated magnetic proximity effect and the formation of moiré miniband is not changed.

In terms of this moiré potential, the Hamiltonian of the heterobilayer can be written as $H = H_0 + V_\tau(\mathbf{R})$, with $H_0$ the original Hamiltonian of monolayer BAs and $V_s(\mathbf{R})$ the moiré potential term. The tight-binding Hamiltonian is

$$H = \sum_{i,\tau}(\varepsilon_i + V_{i,\tau})a_{i,\tau}^\dagger a_{i,\tau} + t\sum_{\langle i,j\rangle,\tau} a_{i,\tau}^\dagger a_{j,\tau} + h.c.$$

where $\langle i,j\rangle$ sums over nearest neighbor pairs with hopping energy $t$ and $\varepsilon_i$ is the onsite energy (see Appendix B). $V_{i,\tau}$ is a spin dependent potential at $i$-th site taking the formula of Eq. (2). Here we focus on the valence band edge where the moiré potential is derived.

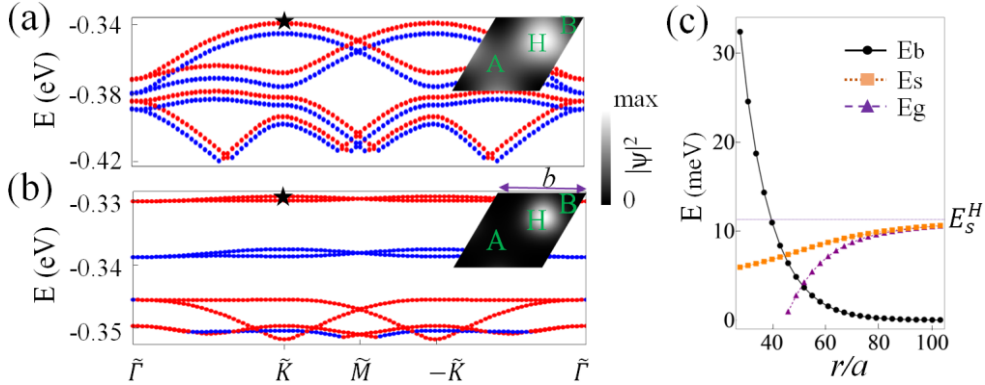

FIG. 5. (a,b) Band structure in the mini-Brillouin zone. The red and blue dots stand for spin-up and spin-down components respectively. The insets show the density distribution of states at $\widetilde{K}$ point for the topmost miniband denoted by the star symbols. The moiré periodicity used is $b = 28a$ for (a) and $b = 67a$ for (b) respectively. (c) Band width $E_b$ (black-solid line with dots) of the top-most miniband, spin splitting Es at $\widetilde{K}$ point (orange-dotted line with squares), and global gap Eg (purple-dashed line with triangles) between top-most spin-up and spin-down minibands as a function of moiré periodicity $b$. The parameters used for monolayer BAs are given in Appendix B. The moiré potential is given in Eq. (2).

The magnetic proximity effect in a moiré is reflected in the spin polarization in the moiré miniband, as shown in Fig. 5. The K and –K valleys in the original Brillouin zone are now folded back to the mini-Brillouin zone and cross at $\widetilde{M}$ point, which is almost degenerate because the intervalley scattering is negligible for a smooth moiré potential. At $\widetilde{\Gamma}$ point, energy gap opens due to the intravalley scattering arising from the moiré modulation. The inset shows distribution of the wave function of the band-edge state, which now becomes localized around the moiré confinement center, located at H stacking region. With the increase of moiré periodicity, the miniband becomes flat and the two spin species are separated (c.f. Fig. 5(b)). Accordingly, the band-edge state becomes more localized.

The formed miniband can be understood from the inter moiré hopping of the localized state, which forms a hexagonal lattice with the same periodicity as moiré superlattice. With the increase of moiré periodicity, the overlap between these states decreases. Accordingly the inter moiré hopping decreases and the miniband become flat. Fig. 5(c) shows that the band width (black-solid line with dots) decays exponentially as a function of moiré periodicity.

In lattice-matched commensurate case, the difference between band-edge energies of the two spins (magnetic proximity effect) also depends on the atomic registry, so there is a modulation of spin splitting in the moiré. This modulation results in the dependence of spin splitting in the miniband on the moiré periodicity and the position where the state is localized. For a short-period moiré, the state is extended over a large area in the superlattice, so the spin splitting in the miniband is almost averaged over all configurations and is small. While for a long-period moiré, the state is localized around some certain configuration with a given atomic registry. Because this area can be approximately described by a lattice-matched commensurate bilayer (H configuration here), the spin splitting value in the miniband will gradually approach to the value at this configuration ($E_s^H$), as the moiré periodicity increases (c.f. orange-dotted line with squares in Fig. 5(c)). Furthermore, for a long-period moiré, there appears a global gap separating the two spin species. The gap (c.f. purple-dashed line with triangles in Fig. 5(c)) approaches the spin splitting value as the miniband become flat.

**B. Electric field controlled magnetic proximity effect**

Beside the electronic property, the interlayer distance also changes smoothly in the moiré. In the moiré pattern formed in MoS$_2$/WSe$_2$ heterobilayer, the STM measured variation of interlayer distance is about 1.8Å [20]. In the presence of a perpendicular electric field, the modulation of interlayer distance results in a modulation of Stark shift energy, which acts as a spin-independent scalar moiré potential. This modulation further reshapes the band-edge energy profile, affecting the localization of the band-edge state and providing a gate controllable way to manipulate the magnetic proximity effect in the moiré.

The atomic registry dependence of the variation of interlayer distance can be formulated as $\delta d(\mathbf{r}) = d_0 \sum_{i=1,2,3} \cos(\mathbf{G}_i \cdot \mathbf{r} + \varphi_d)$, with $\{d_0, \varphi_d\} \approx \{0.12\text{Å}, 167°\}$ being fitted from first-principles result in lattice-matched heterobilayer BAs/CrI$_3$ (c.f. Fig. 6(a)). We assume that the ferromagnetic layer is deposited on a substrate so that there is no out-of-plane corrugation. Then for a perpendicular electric field $\mathcal{E}$, the modulation of Stark shift energy is

$$V_{ext}(\mathbf{r}) = v_0 \sum_{i=1,2,3} \cos(\mathbf{G}_i \cdot \mathbf{r} + \varphi_d)$$

with modulation amplitude $v_0 = e\mathcal{E}d_0$. For electric field with $\mathcal{E}<0$, the confinement center of this energy is located at A configuration.

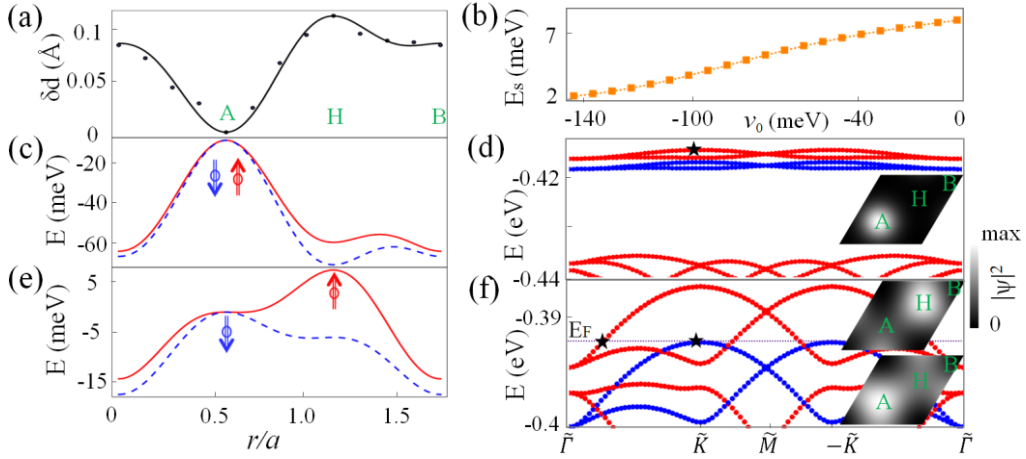

FIG. 6. (a) Interlayer translation dependence of the variation of interlayer distance (relative to A configuration). The relaxed Dots are first-principles result. (b) Spin splitting energy in the miniband as a function of the amplitude of gate controlled Stark shift energy $v_0$. (c,d) Band-edge energy profile and moiré miniband for $v_0 = -140$meV. Inset in (d) shows the density distribution of state at $\widetilde{K}$ point for topmost miniband. (e,f) Energy profile and moiré miniband for $v_0 = -77$meV. The insets in (f) show the density distribution of the spin up (upper inset) and spin down (lower inset) states at energy E$_F$. $b=58a$ is used.

The competition of this gate controlled modulation of Stark shift energy $V_{ext}(\mathbf{r})$ with moiré potential $V_\tau(\mathbf{r})$ defined in Eq. (2) determines the band-edge energy profile in the moiré. The change of this profile affects the localization of band-edge state and hence the magnetic proximity effect in vdW moiré. Fig. 6(b) shows the spin splitting in the moiré miniband as a function of the amplitude of the Stark shift energy $v_0$, showing a monotone decrease as $v_0$ decreases. This is understandable because, with the decrease of $v_0$, the confinement center is changed from H to A local (c.f. Figs. 6(c) and 6(d)). And the spin splitting in A configuration is much smaller compared with the one in H configuration (c.f. Fig. 3(b)).

The spatial modulation of magnetic proximity effect in the moiré provides a possibility to spatially separate carries with different spins. This is achieved when the centers of moiré modulation for the two spins are gate tuned to different locations (c.f. Fig. 6(e)). Fig. 6(f) shows the band structure in this case, with the two spin states now being separated at H and A locals. This spin separation effect arising from the moiré modulation of magnetic proximity effect is unique in vdW heterostructure.

## V. Discussion and summery

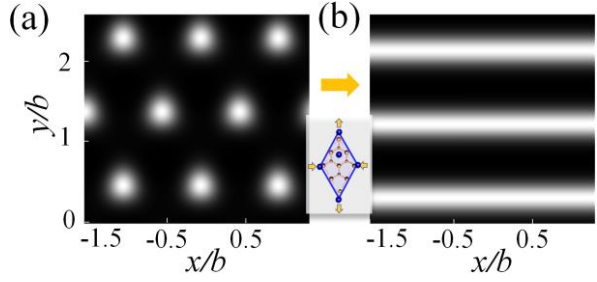

FIG. 7. (a) A schematic of spin-polarized quantum dot arrays in a hexagonal superlattice. The spin-polarized carries are confined at the bright areas. (b) A uniaxial strain (inset) transforms the hexagonal superlattice into a one dimensional superlattice, which serves as waveguide arrays for the spin-polarized carries.

The magnetic proximity effect in vdW moire may lead to opportunities to engineer programmable nanodevices with spin functionality. The moiré potential can serve as the spin-dependent confinement potential, giving rise to spin-polarized quantum dot array (c.f. Fig. 7(a)). The energy levels in the quantum dots can be controlled by changing the moiré periodicity, which is tunable via a relative twisting and/or strain between the layers. More interestingly, applying a uniaxial strain, when the lattice of the bilayer is matched in one direction, the hexagonal superlattice is transformed into a one dimensional superlattice [26]. This one dimensional moiré can define an array of waveguides for transport of spin-polarized carries (c.f. Fig. 7(b)). The direction of proximity-induced spin polarization in the monolayer semiconductor depends on the magnetization in the monolayer FM layer, which can be switched by an external magnetic field. For $CrI_3$, a small magnetic field (on the order of $\sim 10^{-1}$T) is needed to reverse the direction of its magnetization [2].

VdW heterostructures with proximity-induced spin-polarized mini-bands can be used to engineer spin filters and spin memories with high tunability. The amplitude of the spin-polarized current transmitted in the moiré pattern, related to the hopping between neighboring moiré sites, can be controlled by magnetic, electrical and mechanical means. For moiré pattern with large periodicity, the spin polarized states are so localized that inter moire hopping is suppressed. In this case, these well localized states can be used to store spin information, functioning as spin memories. In one dimensional moiré, the spin filter becomes highly anisotropic, and the transmission of spin-polarized current is allowed in the direction with identical lattice constant. More interestingly, this uniaxial spin filter only transmits current at certain regions (c.f. Fig. 7(b)), which forms a one dimensional nanoscaled spin filter array.

In summary, we have studied the magnetic proximity effect in a heterobilayer formed by a monolayer semiconductor on a 2D ferromagnet. The proximity effect stems from the spin dependent interlayer hopping between the layers, which depends sensitively on the interlayer atomic registry. Taking into account the effect of forming moiré pattern due to lattice mismatch or twisting between the layers, we also studied the moiré miniband with band-edge state localized periodically in the heterobilayer. The magnetic proximity effect in the moiré pattern is reflected in the spin splitting in the miniband, which depends on the moiré periodicity. The moiré miniband becomes flat with the increase of moiré periodicity, which provides a platform for studying strong correlation effect. An external electric field tunes the magnetic proximity effect and makes it possible to spatially separate carries with different spins in the moiré. Some programmable nanodevices are envisioned utilizing the magnetic proximity effect in vdW moiré.

**Appendix A. First-principles calculations**

DFT calculations of BAs/CrI$_3$ were performed using the Vienna ab initio Simulation Package [46,47]. The heterostructure is modeled in terms of a slab consisting of a 2x2 supercell of BAs on a 1x1 unit cell of CrI$_3$, which results in a lattice mismatch of 1.3% (lattice constants used are 3.39 Å and 6.867 Å for BAs and CrI$_3$ respectively). In the study of translation dependence, the in-plain positions are kept fixed and the out-of-plane positions are relaxed. The relaxed interlayer distances are 3.78 Å, 3.85 Å and 3.90 Å for A, B and H configurations respectively. To avoid artificial interactions between the polar slabs, two such slabs, oppositely oriented with mirror symmetry, are placed in each supercell, which is separated from its periodic images by 10 Å vacuum regions. The using of mirror-symmetric slabs gives a symmetric electrostatic potential and well defines the vacuum level. The exchange correlation functional is approximated by the generalized gradient approximation as parametrized by Perdew, Burke and Ernzerhof [48], and pseudopotentials were constructed by the projector augmented wave method [49,50]. The 2D Brillouin zone is sampled by a 15x15x1 Monkhorst-Pack mesh. To eliminate band folding caused by the use of supercells the wave functions were projected onto the corresponding momentum vector k of the primitive cell [51,52]. Bands for BAs are then obtained by weighting the bands of the supercell by the contributions of the k-projected wave functions.

**Appendix B. Band structures of monolayer BAs and CrI$_3$**

BAs has planar honeycomb atomic structure as Graphene. The conduction and valence bands have degenerate energy extrema around $\pm K$ points known as valleys. The valence band maximum and conduction band minimum are mainly attributed to the 4$p_z$ orbital of As atom and the 2$p_z$ orbital of B atom respectively.

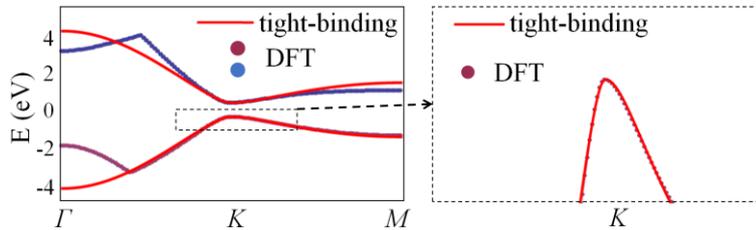

Fig. 8. Band structure of BAs. The dots are first-principles results and the red lines are tight-binding results with fitting parameters $\varepsilon_A = -\varepsilon_B = 0.38$eV and $t$=1.43eV. The right panel shows the band structure around the valence band edge.

As in gaped Graphene, we use the $p_z$ orbits to construct a tight-binding model,

$$H_0 = \sum_i \varepsilon_i a_i^\dagger a_i + t \sum_{<i,j>} a_i^\dagger a_j + h.c.$$

We only consider nearest neighbor hopping here. From fitting to DFT results, we obtain $\varepsilon_A = -\varepsilon_B = 0.38$eV, $t=1.43$eV (c.f. Fig. 8).

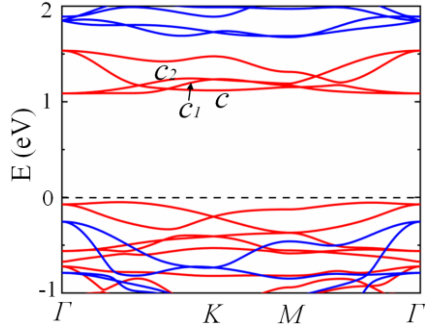

Fig. 9. First-principles band structure of $CrI_3$. The red and blue lines are two spin components. $c$ and $c_{1,2}$ are the three energy bands used in Table I.

Monolayer $CrI_3$ is a ferromagnet with magnetic moment of $3/2\mu_B$ per Cr atom. Fig. 9 shows the band structure of monolayer $CrI_3$. Three energy bands ($c$ and $c_{1,2}$) contributed most to the spin splitting in the valence band edge of $BAs/CrI_3$ heterobilayer are indicated.

**Appendix C. Effects of interlayer distance, spin-orbital coupling, and vdW functional**

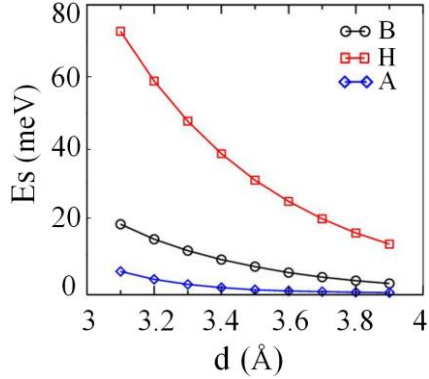

FIG. 10. Spin splitting energy Es as a function of interlayer distance d for the three high-symmetry configurations.

We have performed calculations of the spin splitting as a function of layer distance for the three configurations (c.f. Fig. 10). When reducing the interlayer distance, interlayer coupling increases and the spin splitting increases accordingly. We note that A configuration, which is supposed to have zero spin splitting, also has a small splitting energy when the interlayer distance is substantially reduced. This is because with the interlayer coupling increases greatly, one should take into account contributions from other energy bands, besides the three ones ($c$, $c_1$, and $c_2$ bands in Fig. 9) considered in the main text.

The spin-orbital coupling (SOC) will induce spin-flipped interlayer process. The band splitting induced by this process assisted by SOC is $E'_s \approx \sum_i \left(\frac{1}{E_i+\Delta_i} - \frac{1}{E_i}\right) t'^2_i$, where $t'_i$ is the spin-flipped interlayer hopping amplitude between the layers. This contribution to the splitting has an opposite sign compared with the one from the spin-conserved process. However, when SOC is small, the contribution of the spin-flipped process is negligible, and we only consider the magnetic proximity effect contributed from spin-conserved one.

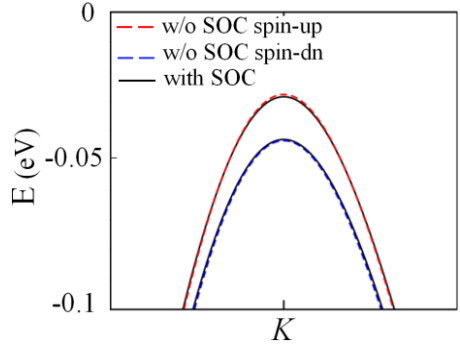

FIG. 11. Band structure around the valence band for H configuration with SOC (black solid lines) and without SOC (red and blue dashed lines).

We have performed SOC calculation for H configuration, in which the spin splitting is most prominent without SOC. From Fig. 11, one can see that the SOC has minor effect on the spin splitting.

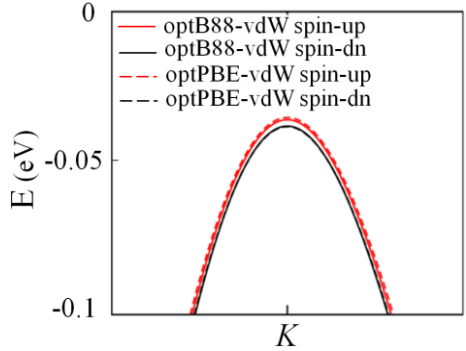

FIG. 12. Band structure around the valence band of B configuration with optB88-vdW functional (solid lines) and with optPBE-vdW functional (dashed lines).

To see how the results depend on the vdW functional, we have performed calculations for B configuration using the optPBE-vdW functional, besides the optB88-vdW functional. We find that there is little difference (less than 0.1 Å) in the equilibrium layer distance between the two functional. From Fig. 12, one can see that the spin splitting is almost unchanged.

**Acknowledgments**

The work is supported by the Croucher Foundation (Croucher Innovation Award), the RGC of HKSAR (HKU17303518, C7036-17W). M C is supported by the National Natural Science Foundation of China (Grants No. 11774084 and No. 91833302). Q T is also supported by the Fundamental Research Funds for the Central Universities from China.

*tongqj@hnu.edu.cn